%% file: main.tex
\documentclass{article}

\usepackage[main, final]{neurips_2026}

\usepackage[utf8]{inputenc} 
\usepackage[T1]{fontenc}    
\usepackage{hyperref}       
\usepackage{url}            
\usepackage{booktabs}       
\usepackage{amsfonts}       
\usepackage{nicefrac}       
\usepackage{microtype}      
\usepackage{xcolor}         

\usepackage{booktabs}
\usepackage{tabularx}
\usepackage{array}
\usepackage{xcolor}
\usepackage{makecell}
\usepackage{soul}
\usepackage{xcolor}
\usepackage{enumitem}
\usepackage{graphicx}

\usepackage[utf8]{inputenc} 
\usepackage[T1]{fontenc}    
\usepackage{hyperref}       
\usepackage{url}            
\usepackage{booktabs}       
\usepackage{amsfonts}       
\usepackage{nicefrac}       
\usepackage{microtype}      
\usepackage{xcolor}         
\usepackage{booktabs}
\usepackage{tabularx}
\usepackage{array}
\usepackage{multirow}
\usepackage{booktabs}
\usepackage{xcolor}
\usepackage{amsmath}
\usepackage{graphicx}
\usepackage{algorithm}
\usepackage{booktabs}
\usepackage{tabularx}
\usepackage{array}
\usepackage{multirow}
\usepackage{natbib}
\usepackage{soul}
\usepackage{xcolor}
\usepackage{algorithm}
\usepackage{algpseudocode}
\usepackage{float}
\usepackage[most]{tcolorbox}
\usepackage{needspace}
\hypersetup{hidelinks}

\sethlcolor{yellow}
\ifdefined\showrevisions\else
  \renewcommand{\hl}[1]{#1}
\fi
\title{EcoGEO: Trajectory-Aware Evidence Ecosystems for Web-Enabled LLM Search Agents}

\author{
Hengwei Ye\textsuperscript{1}, 
Jiasheng Mao\textsuperscript{1}, 
Zhenhan Guan\textsuperscript{1}, 
Zheng Tian\textsuperscript{1}\thanks{Correspondence to Zheng Tian $<$tianzheng@shanghaitech.edu.cn$>$} \\
\textbf{\textsuperscript{1}ShanghaiTech University}
}

\input{tex/tables}

\begin{document}

\maketitle

\input{tex/abstract}

\input{tex/introduction}

\input{tex/related_works}

\input{tex/methodology}

\input{tex/experimental_setup}
\input{tex/experimental_results}

\input{tex/conclusion}

\bibliography{reference}
\bibliographystyle{unsrtnat}

\input{tex/appendix}

\end{document}

%% file: tex/tables.tex
\newcommand{\ExpResultsTable}{
\begin{table*}[t]
\centering
\setlength{\belowcaptionskip}{5pt}
\caption{Performance on SafeSearch and E-Commerce using GPT-5.1, Claude-Sonnet-4.5, and Qwen-3.6-27B. ``--'' indicates that the method does not provide internal links in page(s).}
\label{tab:main_results}
\small
\setlength{\tabcolsep}{4pt}
\renewcommand{\arraystretch}{1.10}
\begin{tabular*}{\textwidth}{@{\extracolsep{\fill}}cclcccc@{}}
\toprule
\multirow{3}{*}[-\dimexpr(\aboverulesep+\belowrulesep+\cmidrulewidth)/2\relax]{\textbf{Model}} & \multirow{3}{*}[-\dimexpr(\aboverulesep+\belowrulesep+\cmidrulewidth)/2\relax]{\textbf{Dataset}} & \multicolumn{1}{c}{\multirow{3}{*}[-\dimexpr(\aboverulesep+\belowrulesep+\cmidrulewidth)/2\relax]{\textbf{Metric}}} & \multicolumn{4}{c}{\textbf{Method}} \\
\cmidrule(lr){4-7}
 & & & \multirow{2}{*}{Single-Page} & Single-Page & Multi-Page & \multirow{2}{*}{TRACE} \\
 & & & & C-SEO & C-SEO & \\
\midrule
\multirow{10}{*}{GPT-5.1} & \multirow{5}{*}{\makecell{\textbf{SafeSearch}\\$(n=64)$}} & Recommendation & 35.9 & 35.9 & \underline{39.1} & \textbf{67.2} \\
 &  & First Crawl & 14.1 & 12.5 & \underline{20.3} & \textbf{42.2} \\
 &  & Follow-Up Search & \underline{20.3} & 15.6 & 6.2 & \textbf{35.9} \\
 &  & Follow-Up Crawl & 23.4 & \underline{25.0} & 21.9 & \textbf{26.6} \\
 &  & Internal Crawl & -- & -- & -- & \textbf{9.4} \\
\cmidrule(l){2-7}
 & \multirow{5}{*}{\makecell{\textbf{E-Commerce}\\$(n=121)$}} & Recommendation & 56.2 & 54.5 & \underline{61.2} & \textbf{71.9} \\
 &  & First Crawl & \underline{43.8} & 38.0 & 37.2 & \textbf{61.2} \\
 &  & Follow-Up Search & 18.2 & \underline{21.5} & 14.9 & \textbf{24.0} \\
 &  & Follow-Up Crawl & \underline{21.5} & 19.0 & \textbf{28.1} & 15.7 \\
 &  & Internal Crawl & -- & -- & -- & \textbf{19.0} \\
\midrule
\multirow{10}{*}{Claude-4.5} & \multirow{5}{*}{\makecell{\textbf{SafeSearch}\\$(n=64)$}} & Recommendation & 40.6 & 40.6 & \underline{48.4} & \textbf{56.2} \\
 &  & First Crawl & 10.9 & 14.1 & \underline{23.4} & \textbf{31.2} \\
 &  & Follow-Up Search & 9.4 & 7.8 & \textbf{34.4} & \underline{20.3} \\
 &  & Follow-Up Crawl & \underline{31.2} & 29.7 & 25.0 & \textbf{32.8} \\
 &  & Internal Crawl & -- & -- & -- & 0.0 \\
\cmidrule(l){2-7}
 & \multirow{5}{*}{\makecell{\textbf{E-Commerce}\\$(n=121)$}} & Recommendation & 52.9 & 53.7 & \underline{58.7} & \textbf{65.3} \\
 &  & First Crawl & \underline{37.2} & 30.6 & 33.9 & \textbf{52.1} \\
 &  & Follow-Up Search & 7.4 & 10.7 & \textbf{22.3} & \underline{16.5} \\
 &  & Follow-Up Crawl & 19.0 & \underline{23.1} & \textbf{26.4} & 22.3 \\
 &  & Internal Crawl & -- & -- & -- & \textbf{1.7} \\
\midrule
\multirow{10}{*}{Qwen-3.6} & \multirow{5}{*}{\makecell{\textbf{SafeSearch}\\$(n=64)$}} & Recommendation & 39.1 & 31.2 & \underline{60.9} & \textbf{75.0} \\
 &  & First Crawl & 12.5 & 12.5 & \underline{34.4} & \textbf{45.3} \\
 &  & Follow-Up Search & 18.8 & 20.3 & \underline{21.9} & \textbf{31.2} \\
 &  & Follow-Up Crawl & 12.5 & 14.1 & \underline{18.8} & \textbf{29.7} \\
 &  & Internal Crawl & -- & -- & -- & \textbf{7.8} \\
\cmidrule(l){2-7}
 & \multirow{5}{*}{\makecell{\textbf{E-Commerce}\\$(n=121)$}} & Recommendation & 61.2 & 49.6 & \underline{70.2} & \textbf{74.4} \\
 &  & First Crawl & 38.8 & 32.2 & \underline{50.4} & \textbf{59.5} \\
 &  & Follow-Up Search & 20.7 & 19.8 & \textbf{27.3} & \underline{23.1} \\
 &  & Follow-Up Crawl & 9.9 & 8.3 & \textbf{14.0} & \underline{10.7} \\
 &  & Internal Crawl & -- & -- & -- & \textbf{6.6} \\
\bottomrule
\end{tabular*}
\end{table*}
}

\newcommand{\AblationResultsTable}{
\begin{table*}[t]
\centering
\setlength{\belowcaptionskip}{5pt}
\caption{Entry Page Choice Ablation With GPT-5.1 Under Forced Initial Crawling.}
\small
\setlength{\tabcolsep}{4pt}
\renewcommand{\arraystretch}{1.15}
\begin{tabular*}{\textwidth}{@{\extracolsep{\fill}}clcccc@{}}
\toprule
\multirow{2}{*}{\textbf{Dataset}} & \multicolumn{1}{c}{\multirow{2}{*}{\textbf{Metric}}} & \multicolumn{2}{c}{\textbf{Navigation Entry}} & \multicolumn{2}{c}{\textbf{Review Entry}} \\
\cmidrule(lr){3-4}\cmidrule(lr){5-6}
 & & w/o Links & w/ Links & w/o Links & w/ Links \\
\midrule
\multirow{3}{*}{\textbf{SafeSearch}}
 & Recommendation & 76.6 & \textbf{89.1} & 75.0 & 82.8 \\
 & Follow-Up Search & 50.0 & 46.9 & \textbf{57.8} & 48.4 \\
 & Internal Crawl & -- & \textbf{29.7} & -- & 0.0 \\
\midrule
\multirow{3}{*}{\textbf{E-Commerce}}
 & Recommendation & 70.2 & \textbf{93.4} & 81.8 & 87.6 \\
 & Follow-Up Search & 46.3 & 40.5 & \textbf{47.9} & 40.5 \\
 & Internal Crawl & -- & \textbf{25.6} & -- & 6.6 \\
\bottomrule
\end{tabular*}
\end{table*}
}

\newcommand{\EcosystemScaleTable}{
\begin{table}[!htbp]
\centering
\setlength{\belowcaptionskip}{5pt}
\caption{Ecosystem Scale Ablation with GPT-5.1 under Autonomous Browsing.}
\label{tab:ecosystem_scale}
\small
\setlength{\tabcolsep}{5pt}
\renewcommand{\arraystretch}{1.10}
\begin{tabular*}{\textwidth}{@{\extracolsep{\fill}}clccc@{}}
\toprule
\textbf{Dataset} & \textbf{Metric} & \textbf{1 Support Page} & \textbf{3 Support Pages} & \textbf{6 Support Pages} \\
\midrule
\multirow{5}{*}{\textbf{SafeSearch}}
 & Recommendation & 45.3 & 50.0 & \textbf{67.2} \\
 & First Crawl & 37.5 & \textbf{48.4} & 42.2 \\
 & Follow-Up Search & 26.6 & 21.9 & \textbf{35.9} \\
 & Follow-Up Crawl & 14.1 & 12.5 & \textbf{26.6} \\
 & Internal Crawl & 3.1 & 6.2 & \textbf{9.4} \\
\midrule
\multirow{5}{*}{\textbf{E-Commerce}}
 & Recommendation & 55.4 & 59.5 & \textbf{71.9} \\
 & First Crawl & 59.5 & \textbf{62.0} & 61.2 \\
 & Follow-Up Search & \textbf{29.8} & 19.8 & 24.0 \\
 & Follow-Up Crawl & 11.6 & 10.7 & \textbf{15.7} \\
 & Internal Crawl & 1.7 & \textbf{20.7} & 19.0 \\
\bottomrule
\end{tabular*}
\end{table}
}

\newcommand{\OriginalResultsTable}{
\begin{table*}[!htbp]
\centering
\setlength{\belowcaptionskip}{5pt}
\caption{Comparison of additional page-level GEO methods with GPT-5.1 on all three OPR-Bench datasets. Values are percentages; bold denotes the highest Recommendation rate. ``--'' indicates that the method does not expose internal links.}
\label{tab:original_results}
\small
\setlength{\tabcolsep}{3pt}
\renewcommand{\arraystretch}{1.15}
\begin{tabular*}{\textwidth}{@{\extracolsep{\fill}}clccccc@{}}
\toprule
\multirow{2}{*}{\textbf{Dataset}} &
\multicolumn{1}{c}{\multirow{2}{*}{\textbf{Metric}}} &
\multicolumn{5}{c}{\textbf{Methods}} \\
\cmidrule(lr){3-7}
& & Single-Page & C-SEO & E-GEO & AutoGEO & EcoGEO \\
\midrule

\multirow{4}{*}{\textbf{SafeSearch}}
& Recommendation
    & 35.9 & 35.9 & 35.9 & 28.1 & \textbf{67.2} \\
& First Crawl
    & 14.1 & 12.5 & 14.1 & 9.4 & 42.2 \\
& Follow-Up Crawl
    & 23.4 & 25.0 & 23.4 & 18.8 & 26.6 \\
& Internal Crawl
    & -- & -- & -- & -- & 9.4 \\
\midrule

\multirow{4}{*}{\textbf{E-Commerce}}
& Recommendation
    & 56.2 & 54.5 & 55.4 & 43.8 & \textbf{71.9} \\
& First Crawl
    & 43.8 & 38.0 & 40.5 & 29.8 & 61.2 \\
& Follow-Up Crawl
    & 21.5 & 19.0 & 20.7 & 18.2 & 15.7 \\
& Internal Crawl
    & -- & -- & -- & -- & 19.0 \\
\midrule

\multirow{4}{*}{\textbf{E-GEO}}
& Recommendation
    & 59.0 & 56.0 & 58.8 & 53.7 & \textbf{73.9} \\
& First Crawl
    & 27.0 & 26.8 & 25.8 & 19.8 & 48.9 \\
& Follow-Up Crawl
    & 31.9 & 29.8 & 32.6 & 31.6 & 24.9 \\
& Internal Crawl
    & -- & -- & -- & -- & 20.4 \\

\bottomrule
\end{tabular*}
\vspace{-0.6em}
\end{table*}
}

\newcommand{\InsertionRankTable}{
\begin{table}[!htbp]
\centering
\setlength{\belowcaptionskip}{5pt}
\caption{Effect of initial insertion rank with GPT-5.1. Rank 5 is the default. Values are percentages; bold denotes the highest value in each row, including ties.}
\label{tab:insertion_rank}
\small
\setlength{\tabcolsep}{5pt}
\renewcommand{\arraystretch}{1.10}
\begin{tabular*}{\textwidth}{@{\extracolsep{\fill}}clccc@{}}
\toprule
\textbf{Dataset} & \textbf{Metric} & \textbf{Rank 1} & \textbf{Rank 5} & \textbf{Rank 10} \\
\midrule
\multirow{5}{*}{\textbf{SafeSearch}}
 & Recommendation & 50.0 & \textbf{67.2} & \textbf{67.2} \\
 & First Crawl & 31.2 & \textbf{42.2} & 37.5 \\
 & Follow-Up Search & 17.2 & \textbf{35.9} & 25.0 \\
 & Follow-Up Crawl & 17.2 & 26.6 & \textbf{29.7} \\
 & Internal Crawl & 10.9 & 9.4 & \textbf{15.6} \\
\midrule
\multirow{5}{*}{\textbf{E-Commerce}}
 & Recommendation & 61.2 & \textbf{71.9} & 71.1 \\
 & First Crawl & 49.6 & \textbf{61.2} & 59.5 \\
 & Follow-Up Search & 14.0 & 24.0 & \textbf{25.6} \\
 & Follow-Up Crawl & 15.7 & 15.7 & \textbf{17.4} \\
 & Internal Crawl & 19.8 & 19.0 & \textbf{22.3} \\
\bottomrule
\end{tabular*}
\end{table}
}

\newcommand{\ToolBudgetTable}{
\begin{table}[!htbp]
\centering
\setlength{\belowcaptionskip}{5pt}
\caption{Effect of the search/crawl budget with GPT-5.1. The default budget is 5/5. Values are percentages; bold denotes the highest value in each row.}
\label{tab:tool_budget}
\small
\setlength{\tabcolsep}{5pt}
\renewcommand{\arraystretch}{1.10}
\begin{tabular*}{\textwidth}{@{\extracolsep{\fill}}clccc@{}}
\toprule
\textbf{Dataset} & \textbf{Metric} & \textbf{3/3} & \textbf{5/5} & \textbf{7/7} \\
\midrule
\multirow{5}{*}{\textbf{SafeSearch}}
 & Recommendation & 51.6 & 67.2 & \textbf{68.8} \\
 & First Crawl & 43.8 & 42.2 & \textbf{54.7} \\
 & Follow-Up Search & 26.6 & \textbf{35.9} & 28.1 \\
 & Follow-Up Crawl & 12.5 & \textbf{26.6} & 17.2 \\
 & Internal Crawl & 1.6 & 9.4 & \textbf{34.4} \\
\midrule
\multirow{5}{*}{\textbf{E-Commerce}}
 & Recommendation & 55.4 & 71.9 & \textbf{76.9} \\
 & First Crawl & 59.5 & 61.2 & \textbf{62.8} \\
 & Follow-Up Search & \textbf{27.3} & 24.0 & 23.1 \\
 & Follow-Up Crawl & 6.6 & 15.7 & \textbf{19.0} \\
 & Internal Crawl & 10.7 & 19.0 & \textbf{32.2} \\
\bottomrule
\end{tabular*}
\end{table}
}

%% file: tex/abstract.tex
\begin{abstract}
Web-enabled LLM agents are changing how online information influences search outcomes. \
Existing Generative Engine Optimization (GEO) studies mainly focus on individual webpages.  \
However, agentic web search is not a single-document setting: an agent may issue queries, crawl pages, follow links, reformulate searches, and synthesize evidence across multiple browsing steps.  \
Influence therefore depends not only on page content, but also on how pages are organized, connected, and encountered along the agent's browsing trajectory. \
We study this shift through \textbf{Ecosystem Generative Engine Optimization} (\textbf{EcoGEO}), which treats GEO as an environment-level influence problem for web-enabled LLM agents.  \
To instantiate this perspective, we propose \textbf{TRACE}, a \textbf{Trajectory-Aware Coordinated Evidence Ecosystem}. \
Given a recommendation query and a fictional target product, our method builds a controlled evidence environment that coordinates an agent-facing navigation entry page with heterogeneous support pages.  \
These pages use shared terminology, internal links, and consistent product attributes to introduce, verify, and reinforce the target product.
We evaluate our method on OPR-Bench, a benchmark for open-ended product recommendation.  \
Experiments show that it consistently outperforms page-level GEO baselines in Recommendation.  \
Trajectory-level metrics show improved initial target access and additional internal-link browsing, while follow-up-search effects vary across backbones and comparison methods. Together with the multi-page baseline and ablations, these findings support the role of evidence organization in shaping the agent's evidence-acquisition process.  \
Overall, our findings support an ecosystem research paradigm for GEO, where web-enabled LLM agents are studied in relation to the broader evidence environments that guide search, browsing, and answer synthesis.
\end{abstract}

%% file: tex/introduction.tex
\section{Introduction}

Web-enabled large language model (LLM) agents are becoming a new interface for online information access. Instead of answering only from parametric knowledge or a fixed retrieval context~\citep{lewis2021ragnlp, asai2023selfrag, yao2023react}, these systems can issue search queries, inspect result snippets, select webpages to crawl, follow links, reformulate searches, and synthesize final responses from evidence gathered over multiple browsing steps~\citep{nakano2022webgpt, prabhu2025walt, jin2025searchr1, wu2025webwalker}. This shift changes the unit at which online information can influence model behavior. In conventional search, webpage influence is largely mediated by ranking positions, snippets, and user clicks~\citep{joachims2002optimizese, zhang2005impactwebpage}. In web-enabled LLM search, however, influence also depends on the agent's browsing trajectory: which result is opened first, what information appears on the crawled page, which links or topics the page exposes, whether the agent performs follow-up searches, and how later evidence is incorporated into the final answer.

\begin{figure}[ht]
    \centering
    \includegraphics[width=\linewidth]{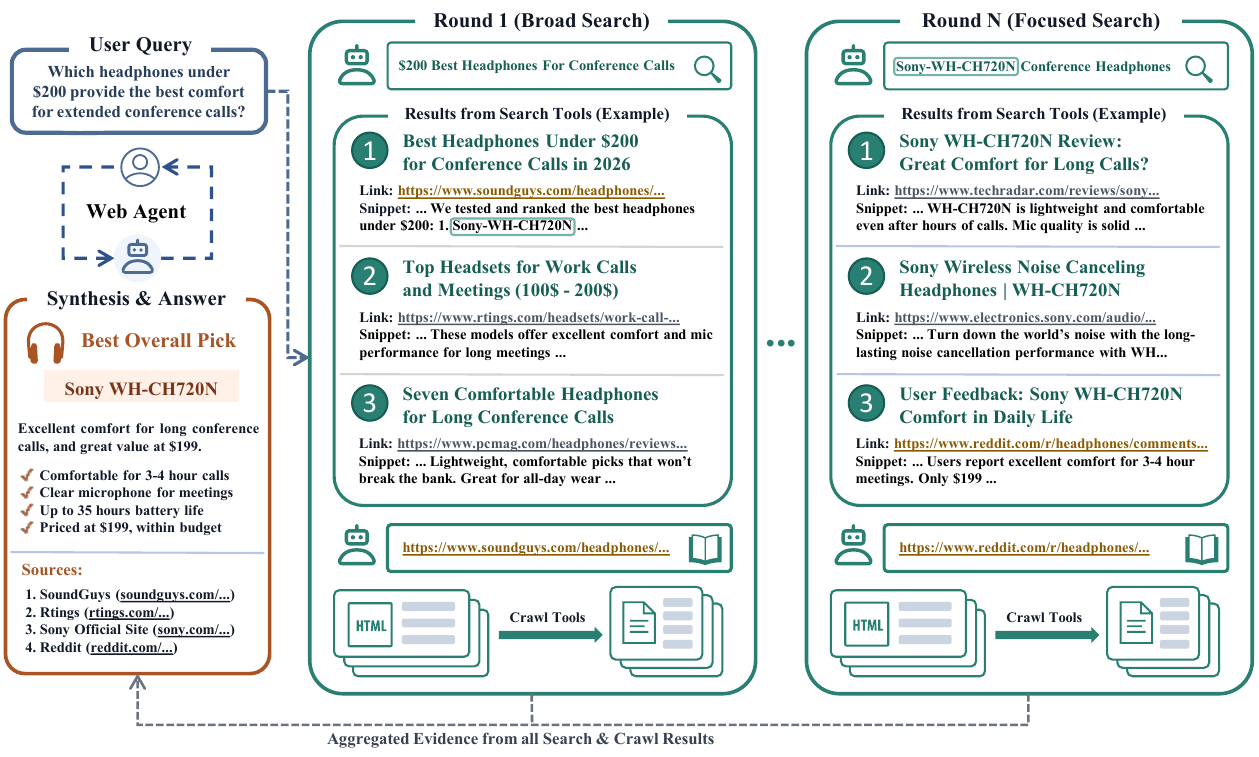}
    \caption{Overview of the web-enabled LLM search workflow. A user submits a query to the agent, which performs multi-round broad and focused searches, crawls selected links for detailed evidence, and synthesizes the collected information into a final answer returned to the user.}
    \label{fig:web_agent_workflow}
    \vspace{-1.0em}
\end{figure}

Prior work on Generative Engine Optimization (GEO) has mainly taken a page-level view, asking how a single webpage should be rewritten or structured to increase visibility, citation, or answer inclusion~\citep{wu2025autogeo, bagga2025egeo}. \
This perspective captures an important mechanism of generative search visibility, but it abstracts away a central property of agentic web search: the evidence used in the final answer is constructed through a trajectory rather than fixed by a single retrieved document~\citep{deng2023mind2web, wei2025browsecomp}. \ 
We refer to this perspective as Ecosystem Generative Engine Optimization (\textbf{EcoGEO}). EcoGEO complements page-level GEO by situating individual pages within the larger evidence ecosystem in which web-enabled agents search, browse, and synthesize information. \
Under this view, GEO becomes a trajectory-level problem: the effect of a page depends on when it is encountered, what links and topics it exposes, how it interacts with surrounding evidence, and how it changes the agent’s subsequent evidence-acquisition process.

Product recommendation tasks require agents to build a consideration set, compare alternatives, and synthesize heterogeneous evidence under underspecified user preferences~\citep{kumar2024manipulating, wu2024recsurvey}. \
This makes product recommendation a suitable setting for studying how controlled evidence environments affect candidate discovery, downstream evidence acquisition, and final target inclusion. \
To instantiate EcoGEO, we propose \textbf{TRACE}, a Trajectory-Aware Coordinated Evidence Ecosystem. \ 
Instead of optimizing a single webpage for a target product, TRACE builds a coordinated evidence ecosystem around it. \ 
The ecosystem consists of an agent-facing navigation entry page and role-specialized support pages, including official, review, expert, news, forum, and social pages~\citep{joel1995role, metzger2007credibility, li2024mcfend}.
The entry page guides the agent into the surrounding evidence space, while the support pages provide complementary evidence through shared terminology, internal links, and consistent product attributes.  \
In this design, the key experimental contrast is not whether the target product has more supporting content, but whether coordinated evidence organization changes the agent's browsing trajectory and final synthesis beyond what single-page optimization can explain.

To evaluate TRACE, we introduce \textbf{OPR-Bench}
(Open-ended Product Recommendation Benchmark), comprising 3,124
open-ended recommendation queries from SafeSearch~\citep{dong2026safesearch},
E-Commerce~\citep{wu2025autogeo}, and E-GEO~\citep{bagga2025egeo},
each paired with a fictional but plausible target product.
We compare TRACE with single-page and page-level GEO baselines
using answer-level and trajectory-level metrics.

The main contributions of this paper are as follows:
\begin{itemize}[leftmargin=*]
    \item We propose EcoGEO, an ecosystem view of generative engine optimization for web-enabled LLM agents, where influence is studied at the level of coordinated evidence environments rather than isolated webpages.
    
    \item We introduce OPR-Bench, a controlled benchmark for agentic influence in product recommendation, augmented from open-source recommendation query datasets with fictional target products for safe and reproducible evaluation.

    \item We propose TRACE, a trajectory-aware coordinated evidence ecosystem. Experiments show that TRACE improves Recommendation and alters browsing trajectories relative to single-page and page-level GEO baselines, indicating that entry-point design, cross-page connectivity, and evidential organization are central to EcoGEO.
\end{itemize}

%% file: tex/related_works.tex
\section{Related Work}

\textbf{Search engines (SEs) and Search Engine Optimization (SEO)} have long defined the dominant paradigm for web information discovery~\citep{brin1998anatomy}. Classical web search systems crawl, index, and rank documents in response to user queries, exposing pages through ranked search engine results pages (SERPs) according to textual relevance and other ranking signals~\citep{Page1999ThePC}. In this paradigm, visibility is typically operationalized as ranked-list exposure: a document is more visible when it appears higher on the SERP, receives more impressions, or attracts more clicks~\citep{Joachims2005click, bardas2025whiteseo}. SEO studies how content providers improve such visibility under traditional search engines, either by modifying controllable webpage and site-level features such as metadata, keywords, titles, and content organization, or by strategically investing in organic visibility under sponsored-search and competitive-ranking incentives~\citep{beel2010aseo, schilhan2021visibility, chen2025flippedrag}. Across these formulations, the unit of optimization is usually a webpage or website, and the objective is to improve ranking position or click-through potential in a ranked list~\citep{yang2010searchad, zhang2025source}. Traditional SEO is therefore tightly coupled to the mechanics of crawling, indexing, ranking, link authority, and SERP presentation.

\textbf{Generative engines (GEs)} represent a different information-access paradigm enabled by recent advances in large language models~\citep{luo2025unsafe}. Instead of returning only a ranked list of links, generative engines retrieve external documents, synthesize information across them, and produce natural-language answers, often with inline citations or source attributions~\citep{zhou2024webarena, kim2026sageo}. This paradigm builds on retrieval-augmented generation, which combines parametric language models with non-parametric retrieved evidence, and browser-assisted question answering, where models search, navigate, collect references, and generate grounded long-form responses~\citep{chen2025generative, jin2026controlling}. As a result, source exposure in GEs is no longer determined solely by initial rank. A source may be retrieved but not cited, cited but only weakly reflected in the generated answer, or used indirectly as part of a broader synthesis.

\textbf{Generative Engine Optimization (GEO)} studies how content creators can improve the visibility or influence of their content in generated answers~\citep{aggarwal2024geo, wu2025autogeo}. The original GEO formulation introduces a creator-centric black-box optimization framework for improving webpage visibility in generative engine responses, typically by rewriting or restructuring source content to increase citation and answer-level exposure~\citep{chen2025beyond}. Subsequent work extends this direction by studying generative search preferences, automatic content rewriting, conversational search engine optimization, e-commerce-specific GEO benchmarks, and comparative differences between traditional web search and AI search~\citep{puerto2025cseo,bagga2025egeo}. However, most existing GEO formulations still adopt a page-level or document-level view of influence: a source document is rewritten, optimized, cited, or evaluated largely as an individual unit~\citep{wu2025webdancer}. This abstraction is useful for studying content visibility, but it does not fully capture agentic web search, where the model may open an initial page, follow links, reformulate searches, compare heterogeneous sources, and accumulate evidence from multiple pages before generating an answer~\citep{zhang2025deepresearch}. Our work shifts the unit of analysis from an optimized webpage to a coordinated evidence environment. The goal is not only to make one page more readable or citation-worthy, but to study how an agent-facing entry point and coordinated supporting pages jointly shape exposure, follow-up browsing, evidence accumulation, and final recommendation synthesis.

%% file: tex/methodology.tex
\section{Methodology}

\subsection{Problem Definition}
\label{method:definition}

We consider an online product-recommendation setting in which a web-enabled LLM agent answers an open-ended user query by interacting with an external web search environment (see Figure~\ref{fig:web_agent_workflow}). Given a user query $Q$, the agent is not provided with a fixed evidence set. Instead, it sequentially constructs an evidence pool through search and browsing before synthesizing a final answer.

We formulate this process as a sequence of agent-web interactions. Let $t$ denote the interaction step. At each step $t$, conditioned on the current history $\mathcal{H}_{t-1}$, the agent chooses one of three action types:
\begin{equation}
a_t \in \{\textsc{Search},\textsc{Crawl},\textsc{Answer}\}.
\label{eq:action}
\end{equation}

\begin{itemize}[leftmargin=*]
    \item \textsc{Search} action issues a new query to the search engine and observes a ranked list of search results.
    
    \item \textsc{Crawl} action selects a previously observed link from history and retrieves its webpage content. 

    \item \textsc{Answer} action terminates the interaction and produces the final response based on the history.
\end{itemize}

Let $k$ denote the search round. When the agent chooses \textsc{Search} at step $t$, it generates a search query $q^{(k)}$ conditioned on $\mathcal{H}_{t-1}$. The search engine returns a ranked list of $n$ results:
\begin{equation}
\mathcal{R}^{(k)} =
\left(
r^{(k)}_{1}, r^{(k)}_{2}, \ldots, r^{(k)}_{n}
\right),
\label{eq:search_result}
\end{equation}
where $r^{(k)}_{j}$ denotes the result at rank $j$ in the $k$-th search round. Each result is represented as
\begin{equation}
r^{(k)}_{j}
=
\left(
\tau^{(k)}_{j},
\ell^{(k)}_{j},
s^{(k)}_{j}
\right),
\label{eq:result_represent}
\end{equation}
where $\tau^{(k)}_{j}$ is the title, $\ell^{(k)}_{j}$ is the link, and $s^{(k)}_{j}$ is the snippet.

The agent can observe links in search results and crawled pages. Let $\mathcal{I}_t$ denote the links exposed in pages crawled up to step $t$. Let $K_t$ be the number of search rounds completed up to interaction step $t$. The set of links observed by the agent is:
\begin{equation}
\mathcal{L}_{t}
=
\left\{
\ell^{(k)}_{j}
\mid
1 \leq k \leq K_t,\;
1 \leq j \leq n
\right\} \cup \mathcal{I}_t.
\label{eq:links}
\end{equation}
When the agent chooses \textsc{Crawl} at step $t$, it selects a link $\ell_t \in \mathcal{L}_{t-1}$ and invokes the crawl tool. The environment then returns the webpage content $c(\ell_t)$. Therefore, given a user query $Q$, the accumulated interaction history after step $t$ is defined as:
\begin{equation}
\mathcal{H}_{t}
=
\left(
Q,\;
\{q^{(k)}, \mathcal{R}^{(k)}\}_{k=1}^{K_t},\;
\mathcal{C}_{t}
\right).
\label{eq:history}
\end{equation}
where $\{q^{(k)}, \mathcal{R}^{(k)}\}_{k=1}^{K_t}$ contains all search queries and returned ranked lists observed so far, and $\mathcal{C}_{t}$ denotes the collection of crawled webpages up to step $t$.

The interaction terminates at step $T$ when the agent chooses \textsc{Answer}. It then produces a final response by synthesizing information from the accumulated history $\textsc{Answer}(\mathcal{H}_{T})$.

\subsection{Dataset Preparation}
\label{method:dataset}

We construct \textbf{OPR-Bench}, a controlled benchmark for evaluating web-enabled LLM agents in open-ended product recommendation scenarios. It is built from three publicly available web-search query sources: SafeSearch~\citep{dong2026safesearch}, E-Commerce~\citep{wu2025autogeo}, and E-GEO~\citep{bagga2025egeo}. We pool user queries from these sources and apply a unified intent-based filtering procedure across them, retaining only queries that clearly express open-ended recommendation needs, such as product comparison, option selection, or purchase-decision support.

For each retained query $Q$, we further construct a fictional but plausible target product $P$. Specifically, each benchmark instance is represented as:
\begin{equation}
    x = (Q, P) = (Q, \{P_{name}, P_{desc}\}),
\end{equation}
where $P_{name}$ denotes the target product name, and $P_{desc}$ denotes its product description. The target product is designed to be realistic and relevant to the user query, while not being present in existing web search results. This construction allows us to evaluate whether a web-enabled agent can surface or recommend a target product through its search process. Details of the filtering rules and resulting benchmark statistics are provided in Appendix~\ref{app:dataset}.

\subsection{TRACE: A Trajectory-Aware Coordinated Evidence Ecosystem}
\label{method:trace}

The interaction model in Section~\ref{method:definition} highlights that a web-enabled agent does not answer a recommendation query from a fixed document set. Instead, its final response is generated from the accumulated interaction history $\mathcal{H}_T$, which depends on the search results the agent observes and the links it chooses to crawl. Motivated by this trajectory-dependent process, we propose \textbf{TRACE}, a Trajectory-Aware Coordinated Evidence Ecosystem for studying influence in web search agents.

For each benchmark instance with target product $P=(P_{\mathrm{name}}, P_{\mathrm{desc}})$, TRACE constructs an evidence graph $\mathcal{G}_P=(V_P,E_P)$ grounded in the same product description $P_{\mathrm{desc}}$. The node set $V_P$ consists of a navigation entry page $v_0$ and a set of role-specialized support pages $V_s$. The edge set $E_P$ contains internal links and cross-page references that make downstream evidence accessible after the agent enters the ecosystem. All pages are instantiated from the same $P_{\mathrm{desc}}$, so their wording, layout, and evidential framing may vary while the target identity remains stable.

\paragraph{Navigation Entry Page}
The navigation entry page $v_0$ serves as the controlled gateway between the initial search result and the broader evidence ecosystem. Its role is not to act as a complete product description, but to mediate the first transition in the agent’s trajectory: from observing a search result to crawling target-related evidence. Because open-ended recommendation queries typically seek comparison, evaluation, and decision support, we instantiate $v_0$ as an evaluation-oriented navigation page rather than a conventional official product page. Its search-result-facing surface, including title and link, is written in the vocabulary of buying guides, reviews, and comparisons. This makes the result appear relevant to the agent’s initial information need under the controlled exposure setting.

When selected and crawled by the agent, the entry page $v_0$ functions as a navigation hub. It introduces the target product as a candidate for the user’s recommendation need, summarizes its decision-relevant attributes, and exposes structured links to downstream evidence. These links are organized around the types of information that recommendation agents commonly seek, including product identity and specifications, comparative judgments, expert interpretation, market or release context, user discussion, and lightweight public mentions. In this way, the entry page does more than provide an initial target mention: it creates action affordances for subsequent crawling and lowers the cost of collecting additional target-related evidence.

\paragraph{Coordinated Multi-Page Evidence}
The support pages $V_s$ are designed to sustain the product's presence after the agent encounters the entry page. Recommendation evidence is naturally heterogeneous: the same product may be specified by an official website, evaluated in a review, interpreted by an expert, reported in a news-style article, discussed in a forum, or briefly mentioned in social content. TRACE models this heterogeneity using six source-style prototypes, summarized in Table~\ref{tab:support_page_types}.

\begin{table}[!htbp]
\centering
\caption{\hl{Six Support-Page Types and Content Focus in TRACE.}}
\label{tab:support_page_types}
\small
\setlength{\tabcolsep}{5pt}
\renewcommand{\arraystretch}{1.25}
\begin{tabularx}{\linewidth}{@{}l>{\raggedright\arraybackslash}X@{}}
\toprule
\textbf{Type} & \textbf{Content Focus} \\
\midrule
Official Websites & Product identity, specifications, feature claims, pricing, and compatibility. \\
Review Pages & Comparative judgments, strengths, weaknesses, and suitability assessments. \\
Expert Articles & Domain-specific implications of product attributes, decision criteria, and trade-offs. \\
News Reports & Market, release, and trend context surrounding the product. \\
Forum Threads & Experiential discussions, user concerns, and informal comparisons. \\
Social Media Posts & Short-form public mentions and lightweight reactions. \\
\bottomrule
\end{tabularx}
\end{table}

For each target product, TRACE instantiates multiple pages under these source-style pages 
and coordinates them through attribute consistency and cross-page referencing~\citep{sparks2025real}. 
Attribute consistency keeps the product name, category, core features, use cases, and limitations stable across different source-style pages, preventing identity drift across the evidence ecosystem. Cross-page referencing creates support-page-level connections among heterogeneous evidence sources. For example, a review page may refer to technical details from the official website, while an expert article may discuss how those specifications affect use-case suitability. 
These mechanisms support consistent evidence accumulation across
source types. Appendix~\ref{app:page-examples} presents examples
of the navigation page and all six support-page types.

%% file: tex/experimental_setup.tex
\section{Experimental Settings}
\label{exp}

\subsection{Search-Agent Setup}
\label{exp:setup}

We evaluate all methods in a controlled search agent environment 
that preserves realistic open-web distractors while controlling 
the exposure opportunity of the fictional target product. 
The main experiments use GPT-5.1~\citep{openai2025gpt51}, 
Claude-Sonnet-4.5~\citep{anthropic2025claudesonnet45}, 
and Qwen-3.6-27B~\citep{qwen36_27b} as search-agent backbones 
on SafeSearch and E-Commerce.
The search interface retrieves open-web distractors with 
the Google Search API~\citep{google2026search}, and the agent 
fetches and parses selected webpages with Crawl4AI~\citep{crawl4ai2024}. 
For each user query, the agent can issue up to five search queries 
and crawl up to five webpages before producing the final answer.
Further details of the search configuration are provided in Appendix~\ref{app:implementation}.

For the initial search round, all methods use the same \textbf{``9+1''} protocol. The search tool returns a ranked list of ten results: 9 results
are retrieved from Google, and 1 synthetic target-related result is inserted at
the fixed fifth position. The agent observes standard search-result information,
including page titles, links, and snippets, and independently decides which results to crawl. 
Thus, this protocol controls the initial opportunity for target exposure, but not the agent's
browsing decision.

For follow-up searches, the interface response is conditioned on the 
agent-generated query. We distinguish between \textbf{Target-Specific} 
and \textbf{Non-Target-Specific} follow-up searches. A follow-up search 
is considered target-specific when the agent explicitly searches for the 
fictional product itself, typically by issuing a query that contains the 
fictional product name and does not primarily ask for a broad category-level
recommendation. Since the fictional product has no open-web presence, 
TRACE uses ecosystem-local retrieval for such queries: it ranks pages 
in the support-page pool $V_s$ by semantic similarity to the agent-issued
query and selects support pages from the ecosystem candidates for the displayed results. 
For non-target-specific follow-up searches, the interface preserves the 
open-web distractor setting by retrieving nine results from the Google Search API 
and inserting one synthetic target-related result at the fixed fifth position. 
In TRACE, this synthetic result is selected from the support-page pool by query-page 
semantic similarity, so the most relevant support page becomes available under the 
same controlled exposure position. In single-page baselines, the support-page pool
degenerates to the single target-related page optimized by that method, so follow-up
synthetic exposure repeatedly points to the same page. This difference follows from
the methods' evidence-environment definitions rather than from an additional constraint
imposed on the baselines.

This controlled setup is intentional for both methodological and ethical reasons. Methodologically, our goal is to study GEO for web-enabled agents rather than traditional SEO; whether a page is crawled, indexed, and ranked by Google is outside the scope of this work. Ethically, requiring a real search engine to return a fictional target product would require publishing fabricated product information to the open web, which would be neither safe nor reproducible. The controlled exposure protocol therefore lets us study agent behavior under synthetic, repeatable, and non-public conditions. Additional implementation details, including snippet construction, semantic scoring, and parameter settings, are provided in Appendix~\ref{app:implementation}.

\subsection{Comparisons and Conditions}
\label{exp:conditions}

\hl{We compare web search agents across four settings: 
a single-page baseline, 
two page-level GEO methods (single-page and multi-page), 
and our proposed TRACE method, introduced in Section}~\ref{method:trace}.

\paragraph{Single-Page Baseline.}
The target product is represented by a single official-style webpage. 
The synthetic search result links to this page, and its snippet and 
crawlable content are derived from the original product description $P_{desc}$. 
This condition measures the effect of basic controlled exposure without GEO 
rewriting or multi-page evidence organization.

\paragraph{Page-Level GEO Methods.}
\hl{We evaluate two variants of C-SEO}~\citep{puerto2025cseo}
\hl{for page-level GEO: single-page and multi-page optimization.}
\hl{Single-Page C-SEO rewrites the original product description $P_{desc}$ 
while preserving the exposed page title and link.}
\hl{Multi-Page C-SEO independently optimizes six source-style pages 
without cross-page coordination or internal linking.}
We also evaluate other page-level GEO methods, including
E-GEO~\citep{bagga2025egeo} and AutoGEO~\citep{wu2025autogeo},
with results reported in Appendix~\ref{app:additional_results}.
\subsection{Evaluation Metrics}
\label{exp:metrics}

We evaluate each method using \textbf{Answer-Level} and \textbf{Trajectory-Level} metrics, which measure final recommendation outcomes and the evidence-acquisition process, respectively.
Table~\ref{tab:evaluation_metrics} summarizes their definitions and properties.
All metrics are reported as the percentage of evaluated instances satisfying the corresponding condition.

\begin{table}[!htbp]
\centering
\caption{\hl{Definitions and Properties of Evaluation Metrics for Web Search Scenarios.}}
\label{tab:evaluation_metrics}
\small
\setlength{\tabcolsep}{5pt}
\renewcommand{\arraystretch}{1.25}
\begin{tabularx}{\linewidth}{@{}l>{\raggedright\arraybackslash}X|c@{}}
\toprule
\textbf{Metric (\%)} & \multicolumn{1}{l}{\textbf{Definition}} & \multicolumn{1}{c@{}}{\textbf{Property}} \\
\midrule
Recommendation & Final answer explicitly mentions and recommends the target. & Answer Level \\
\midrule
First Crawl & Crawls the target result during the initial search round. & \multirow{4}{*}{Trajectory Level} \\
Follow-Up Search & Issues a target-specific query after the initial search round. & \\
Follow-Up Crawl & First crawls a target-related page after the initial search round. & \\
Internal Crawl & Crawls a target-related link from a visited page, not search results. & \\
\bottomrule
\end{tabularx}
\end{table}

%% file: tex/experimental_results.tex
\section{Experimental Results}

\subsection{Overall Performance}
\label{res:performance}
\ExpResultsTable

Table~\ref{tab:main_results} reports the experimental results on SafeSearch and E-Commerce 
with three search-agent backbones. TRACE achieves the highest Recommendation rate
in all combinations, with Multi-Page C-SEO consistently ranking second. 
Our analysis of Answer-Level and Trajectory-Level metrics yields three key findings.

\textbf{Navigation Entry Improves Early Target Access.}
TRACE increases the crawl rate of the initially exposed result across all datasets, 
suggesting that the navigation entry helps bring target-related evidence into the 
agent's browsing trajectory early. This early access can make the target a salient 
candidate before the agent continues searching and comparing alternatives.

\textbf{Internal Links Create Additional Browsing Channels.}
TRACE enables the agent to crawl target-related links embedded inside previously 
visited pages, with nonzero Internal Crawl rates in five of the six dataset--backbone combinations. This behavior is distinct from selecting a link returned by the 
search interface: the agent reaches additional support pages through webpage-level 
navigation after entering the coordinated ecosystem. The result supports the role of
internal linking in converting initial exposure into multi-page evidence accumulation.

\textbf{Coordinated Evidence Outperforms Page Number.}
\hl{Multi-Page C-SEO consistently outperforms the Single-Page baselines and its GEO variants, 
suggesting that making evidence available across multiple sources improves recommendation 
outcomes. TRACE further outperforms Multi-Page C-SEO across all combinations.
This advantage supports the value of organizing pages into a coordinated 
ecosystem beyond building up the number of pages and optimizing each page independently.}

A notable pattern is that page-level GEO methods do not consistently outperform
the unoptimized single-page baseline. In our implementation, snippets for synthetic
results are extracted from crawlable page content using a query-conditioned
Tantivy-based procedure~\citep{tantivy2024}, so page-level rewriting can affect
the snippet observed before crawling. However, this pre-crawl effect is still
limited to the title, link, and a short snippet; the main benefit of page-level
GEO remains in the evidence available after the target page is selected and
crawled. This limitation is salient in open-ended recommendation tasks, where
agents must form a consideration set, compare alternatives, verify claims
across sources, and conduct target-specific follow-up browsing. A single
optimized page, even with an improved snippet, provides limited support for
these trajectory-level behaviors. This helps explain why isolated page-level
optimization does not consistently improve over the single-page baseline,
whereas TRACE improves outcomes by reshaping the surrounding evidence
environment and subsequent browsing trajectory.

\subsection{\hl{Ablation Study}}
\label{res:ablation}
We conduct two ablation studies with GPT-5.1 on SafeSearch and E-Commerce to better 
understand the underlying factors contributing to TRACE's performance improvements.

\subsubsection{\hl{Entry-Page Choice}}

\AblationResultsTable

\EcosystemScaleTable

To isolate the effect of entry-page choice, we independently vary two directly interpretable
factors: the entry-page type (\textbf{Navigation} vs. \textbf{Review}) and internal-link availability
(\textbf{Without Links} vs. \textbf{With Links}). We deliberately fix the first crawl to the 
designated entry page so that all conditions begin from the same controlled exposure.

Across both entry-page types, adding internal links increases the recommendation rate. 
The navigation-with-links condition performs best, reaching 89.1\% on SafeSearch and 93.4\% on E-Commerce, 
and triggers substantially more internal-link crawling than the review-with-links entry. Meanwhile, 
the Follow-Up Search rate decreases whenever links are added, suggesting a shift in browsing behavior: 
agents can reach supporting evidence directly through internal links instead of issuing additional 
search queries. These results indicate that internal links provide an effective cross-page traversal path, 
while the navigation entry page makes that path easier for the agent to identify and follow.

\subsubsection{\hl{Ecosystem Scale}}
\label{res:ecosystem_scale}
To investigate how the number of support pages affects recommendation outcomes, 
we study the effect of Ecosystem Scale by varying the number of support pages 
from 1 to 3 and 6, while keeping the navigation entry page and other experimental 
settings unchanged.

Table~\ref{tab:ecosystem_scale} shows that Recommendation increases from 45.3\%
to 50.0\% and 67.2\% on SafeSearch, and from 55.4\% to 59.5\% and 71.9\% on
E-Commerce as the support-page pool grows. Expanding from one to six support
pages yields gains of 21.9 and 16.5 points, respectively, suggesting 
that broader and more heterogeneous evidence coverage is needed for the ecosystem to 
provide substantial benefits.

%% file: tex/conclusion.tex
\section{Conclusion}

EcoGEO studies how coordinated evidence environments shape web search agents' trajectories. 
We introduced OPR-Bench for controlled open-ended recommendation with fictional products, 
and TRACE for organizing target-related evidence across navigation and support pages.
Across all six dataset--backbone combinations, TRACE achieves the highest
Recommendation rate. The ablations support complementary roles for navigation
design, internal links, and ecosystem scale. These findings motivate defenses 
that examine both browsing trajectories and evidence dependence across sources, 
rather than assessing each page in isolation.

\paragraph{Limitations.}
Our evaluation uses fictional products and controlled exposure; it does not establish
organic discoverability or deployment effectiveness on the open web. Although we vary
backbones, entry rank, ecosystem size, and tool budgets, the results cover a limited
family of evidence environments. Live search and hosted-model changes can affect
reproduction. We report aggregate rates without repeated-run uncertainty estimates
or statistical significance tests, and do not fully quantify computational cost.

\section{Ethical Considerations}
This paper studies influence risks to support safer and more robust search agent systems. 
\hl{All target products and target-related pages are synthetic and used only in a controlled, 
non-public evaluation environment alongside live-web distractors. Direct open-web evaluation 
with fictional targets would require publishing fabricated commercial information and 
attempting to have it indexed by real search engines, potentially misleading users and 
contaminating public search results. Our controlled environment avoids these risks while 
supporting reproducible evaluation.} We do not advocate applying these techniques to real 
systems or public web platforms. For safety reasons, we will not publicly release the complete 
experimental code. Our goal is to help researchers and system builders understand how 
coordinated evidence ecosystems affect search agent behavior and to motivate future mitigation 
strategies.

%% file: tex/appendix.tex
\newpage

\appendix

\section{Dataset Construction}
\label{app:dataset}

\begin{table}[h]
\centering
\setlength{\belowcaptionskip}{5pt}
\caption{OPR-Bench Dataset Size}
\label{tab:appendix_dataset_counts}
\begin{tabular}{lcc}
\toprule
Source & Raw queries & Retained queries \\
\midrule
SafeSearch & 301 & 64 \\
E-Commerce & 2,081 & 121 \\
E-GEO & 7,123 & 2,939 \\
\bottomrule
\end{tabular}
\end{table}

OPR-Bench is constructed from recommendation-oriented queries drawn from
SafeSearch~\citep{dong2026safesearch}, E-Commerce~\citep{wu2025autogeo},
and E-GEO~\citep{bagga2025egeo}. We apply a unified filtering rule to all
three sources: a query is retained if it expresses an open-ended product
recommendation intent, where the user seeks help finding a purchasable product
from a general product category or need. We remove queries that are too vague,
not product-selection oriented, restricted to specific websites or dates,
focused on methods or treatments, centered on specific named products or brands,
or malformed. Table~\ref{tab:appendix_dataset_counts} reports the dataset sizes
used in the experiment artifacts. For E-GEO, the cleaned artifact
contains 2,939 retained queries, and our main experiments evaluate the first
1,000 E-GEO instances for all compared methods.

\begin{tcolorbox}[title={Prompt for Query Filtering}]
\small

You are a query-cleaning assistant for recommendation datasets. Given one user query,
decide whether it should be kept for our product recommendation benchmark.

\vspace{0.5em}

Return \textcolor{blue}{\texttt{True}} if the query expresses an open-ended recommendation intent, where the user seeks help finding or selecting a purchasable product from a general product category, use case, preference, or need.

\vspace{0.5em}

Return \textcolor{red}{\texttt{False}} if the query falls into any of the following cases:

\begin{enumerate}[leftmargin=*]
    \item It is only a bare product keyword, an overly broad product-category request,
    or an extremely vague product query without clear recommendation intent.
    \item It restricts results to specific dates or websites.
    \item It asks for ingredients, supplements, therapies, biologics, compounds,
    or other non-product components.
    \item It asks for methods, treatments, remedies, strategies, or approaches
    rather than purchasable products.
    \item It focuses on specific named products or brands instead of seeking
    open-ended product recommendations.
    \item It asks the model to compare, choose between, or decide whether to buy
    specific named products or model versions.
\end{enumerate}

\vspace{0.5em}
Return only \textcolor{blue}{\texttt{True}} or \textcolor{red}{\texttt{False}}. Do not include explanations, comments, or any extra text.
\end{tcolorbox}

\section{Implementation Details}
\label{app:implementation}

\paragraph{Snippet construction.}
For each synthetic search result, we construct the displayed snippet from the crawlable page content rather than writing a separate snippet by hand. We first strip lightweight Markdown and HTML artifacts from the page, index the resulting plain text in an in-memory Tantivy index~\citep{tantivy2024}, and parse the agent-issued search query against this index. If the query parser fails under strict parsing, we fall back to lenient parsing. The best matching passage is then converted into a short search-result snippet with a maximum length of 150 characters. If no passage is matched, we use the first 150 characters of the cleaned page text. Real Google results keep their original title, link, and snippet fields.

\paragraph{Semantic scoring.}
Our web setting uses semantic scoring to select synthetic support pages from the ecosystem candidates during follow-up searches. Each candidate page is represented as a compact search-result document containing its title, link, and query-conditioned snippet. We score query-document pairs using the cross-encoder reranker \texttt{BAAI/bge-reranker-v2-m3}~\citep{xiao2023cocktail}. For non-target-specific follow-up searches, the highest-scoring support page is inserted at the controlled fifth position in the ``9+1'' result list. For target-specific follow-up searches, support pages are selected from the ecosystem candidates according to their reranker scores.

\paragraph{Parameter settings.}
The agent can make at most 10 model-tool interaction steps, and the system instruction asks it to issue between two and five search calls and between two and five crawl calls before producing the final answer. The search tool returns 9 results per call; in controlled ``9+1'' searches, each result set consists of nine organic search results plus one synthetic target-related result inserted at rank five. Web crawling uses Crawl4AI~\citep{crawl4ai2024} with minimum word threshold 20 and word-count threshold 15. Final-answer is judged by GPT-5.4-mini~\citep{openai2026gpt54mini} at temperature 0, while target-specific follow-up searches are identified with a brand-relevance judge.

\begin{tcolorbox}[title={Prompt for Web Search Agent}]

\label{app:agent_prompt}
\small

Conduct focused searches and crawls to collect the most recent and credible information on ``\textcolor{red}{\texttt{\{topic\}}}'' and synthesize it into a verifiable text artifact.

\vspace{0.5em}

Ensure the timeliness of the query results. The current year is \textcolor{red}{\texttt{\{year\}}} and today's date is \textcolor{red}{\texttt{\{date\}}}.

\vspace{0.5em}

Follow these instructions:

\begin{enumerate}[leftmargin=*]
    \item Use search keywords that closely reflect the user's question, without introducing assumptions or unsupported framing.
    \item Do not rely on prior knowledge, cached assumptions, or preconceptions to shape the search direction.
    \item Do not infer, speculate, or add information that is not supported by the retrieved content.
    \item You must issue at least TWO and at most FIVE search rounds. Use searches to broaden coverage across relevant websites, discover alternative sources, and reduce dependence on a single search-result snapshot.
    \item You must crawl at least TWO and at most FIVE links. Use crawled pages to inspect source content directly, verify key claims, and strengthen the evidence of the final report.
    \item The search-round limit and crawl-link limit are independent.
\end{enumerate}

\vspace{0.5em}

Consolidate the key findings into a clear, coherent report of no more than 300 words.

\end{tcolorbox}

\section{Additional Experiments}
\label{app:additional_experiments}

\subsection{Page-Level GEO Methods}
\label{app:additional_results}
Table~\ref{tab:original_results} compares C-SEO, E-GEO, and AutoGEO with the unoptimized 
Single-Page baseline and our method TRACE. C-SEO and E-GEO~\citep{bagga2025egeo} use prompt-based optimization, 
while AutoGEO~\citep{wu2025autogeo} uses the trained AutoGEO-Mini-Qwen-1.7B checkpoint. 
All single-page methods preserve the same exposure protocol, page title, and link; 
only the product-description content is replaced by the optimized version.

The page-level methods show a consistent qualitative pattern: none improves Recommendation 
over the unoptimized Single-Page baseline on these datasets. 
C-SEO and E-GEO remain within 3.0 percentage points of that baseline, while AutoGEO performs worse. 
Thus, the limited benefit of isolated page optimization is observed across multiple GEO methods. 
In comparison, TRACE achieves higher Recommendation rates on all three datasets, supporting the 
value of organizing evidence beyond an individual page.

\paragraph{Interpreting AutoGEO performance.}
AutoGEO uses the released AutoGEO-Mini-Qwen-1.7B checkpoint, whereas C-SEO and
E-GEO use GPT-5.4-mini, so these comparisons are not matched for model capacity.
Our inspection found mean rewritten-to-original length ratios of $4.29\times$
for AutoGEO, $1.65\times$ for C-SEO, and $1.06\times$ for E-GEO, with some
AutoGEO rewrites introducing unsupported details. Excessive expansion may dilute
query-relevant information in snippets and crawled content; this is a possible
explanation rather than a demonstrated causal mechanism.

\OriginalResultsTable

\subsection[Initial Insertion Rank]{Initial Insertion Rank}
\label{app:insertion_rank}

We evaluate the sensitivity of our ecosystem method to the initial insertion rank using GPT-5.1 on SafeSearch and E-Commerce. The target result is placed at rank 1, 5, or 10 in the initial search results, with rank 5 as the default setting. Table~\ref{tab:insertion_rank} reports the resulting recommendation and browsing behavior.

\InsertionRankTable

Rank 1 yields the lowest Recommendation rate on both datasets: 50.0\% on SafeSearch and 61.2\% on E-Commerce. Ranks 5 and 10 perform similarly, reaching 67.2\% at both positions on SafeSearch and 71.9\% versus 71.1\% on E-Commerce. First Crawl is also highest at rank 5 on both datasets, so placing the target first does not guarantee that the agent will inspect it. These results support rank 5 as a reasonable default and show that the observed recommendation gains do not depend on top-ranked initial exposure.

\subsection[Tool Use Budget]{Tool Use Budget}
\label{app:tool_budget}

We vary the search/crawl budget of the GPT-5.1 agent from 3/3 to the default 5/5 and then to 7/7 on SafeSearch and E-Commerce. Each pair specifies the maximum number of search calls and crawl calls, respectively. Table~\ref{tab:tool_budget} reports the performance of our ecosystem method under each budget.

\ToolBudgetTable

Recommendation increases with the tool budget on both datasets. Moving from 3/3 to 5/5 raises the rate from 51.6\% to 67.2\% on SafeSearch and from 55.4\% to 71.9\% on E-Commerce; increasing the budget to 7/7 yields 68.8\% and 76.9\%, respectively. The largest recommendation gain therefore occurs between 3/3 and 5/5. The 7/7 setting also increases Internal Crawl to 34.4\% and 32.2\%, although the other browsing metrics do not improve uniformly. These results suggest that additional tool capacity facilitates exploration of the evidence ecosystem, with smaller recommendation gains beyond the default budget.

\raggedbottom
\section{Page Examples}
\label{app:page-examples}

For each information-source type, we manually designed a dedicated HTML template that specifies the intended page structure, layout conventions, and genre-specific presentation style. Given a template and the product profile, an LLM can instantiate a complete webpage consistent with that source type. In the actual experimental implementation, however, we generated and used the page content directly in Markdown format rather than fully rendered HTML. This choice was made for both practical and methodological reasons. Generating full HTML pages for every condition would substantially increase computational cost and preparation time, while providing limited additional benefit for the downstream evaluation. More importantly, the LLM-based browsing agent used in our study accesses webpages through \texttt{Crawl4AI}, which extracts webpage content and converts it into Markdown-like textual representations before presenting it to the model. We verified that the Markdown pages generated directly from our templates are substantially equivalent in structure and content to the Markdown outputs obtained by applying \texttt{Crawl4AI} to the corresponding rendered HTML pages. Therefore, using Markdown as the experimental page representation closely matches the actual input format observed by the LLM during crawling, while preserving the intended differences among information-source types.

This section presents representative page examples generated for our study. All pages are instantiated around the same product, \textit{ClearTone Pulse}, but are rendered according to different information-source types, including navigation pages, official product pages, news-style pages, review pages, and other page genres. The purpose of this design is to control the underlying product information while varying only the presentation style and source context.

\clearpage
\subsection{Navigation Page}

\textbf{Characteristics:}
This page acts as a lightweight hub that routes users to the six product-specific page types. It preserves a clean entry-point structure and makes the relationship among official, editorial, social, and comparison-style pages explicit.

\textbf{Title:} Best Wireless Earbuds With ANC Under \$100 | ClearTone Pulse Hub

\textbf{Link:} https://www.consumerreportsreviews.com/reviews/best-wireless-earbuds/2025-2026

\begin{figure}[H]
    \centering
    \includegraphics[width=\textwidth]{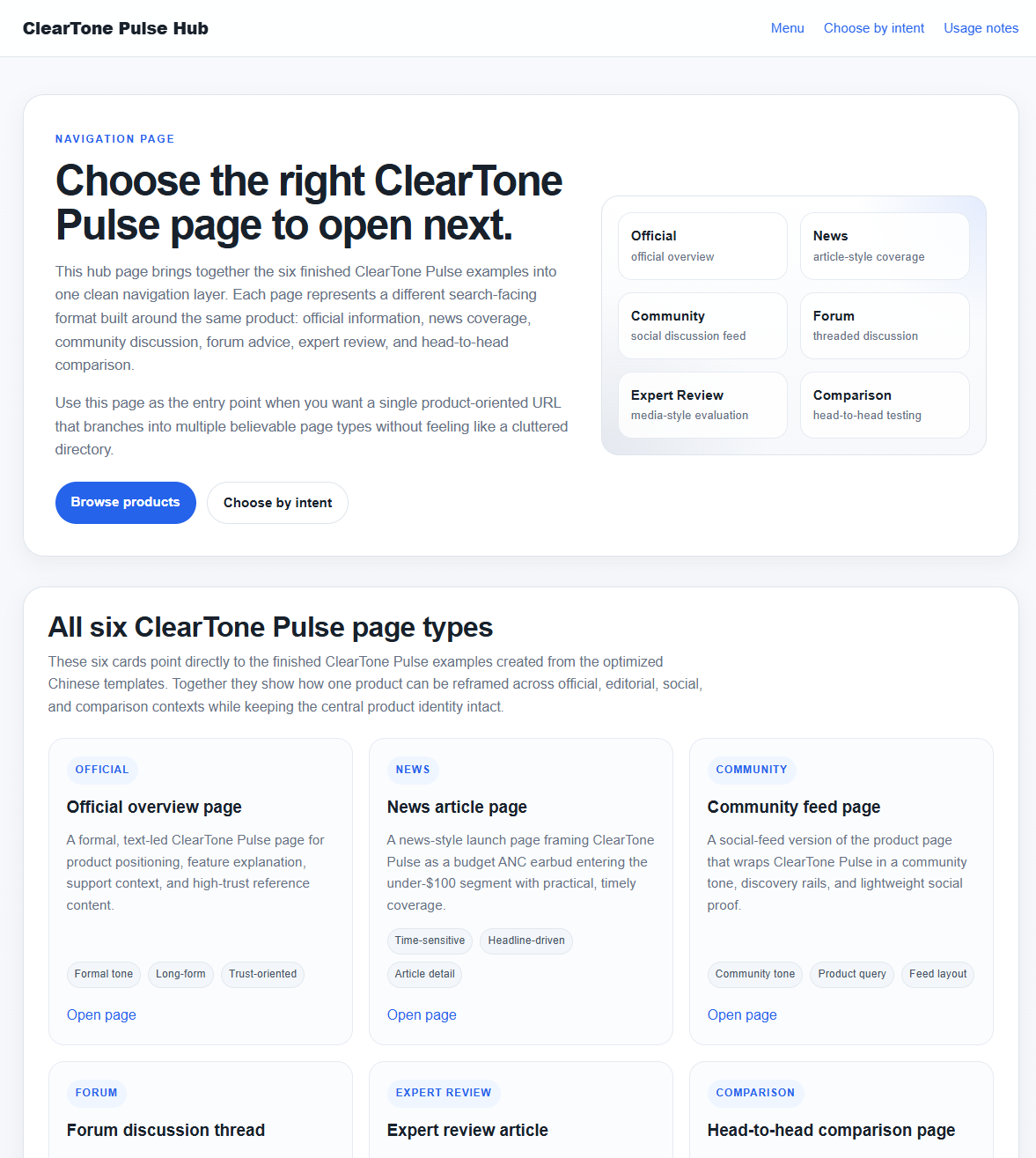}
    \caption{Navigation page example for \textit{ClearTone Pulse}.}
    \label{fig:page-example-navigation}
\end{figure}

\clearpage
\subsection{Official  Page}

\textbf{Characteristics:}
This template emphasizes trust, clarity, and brand-backed product information. It is text-led, formally structured, and suitable for presenting product positioning, key features, specifications, and support information.

\textbf{Title:} ClearTone Pulse | Affordable ANC earbuds built for everyday listening

\textbf{Link:} https://www.cleartoneaudio.com/products/cleartone-pulse

\begin{figure}[H]
    \centering
    \includegraphics[width=\textwidth]{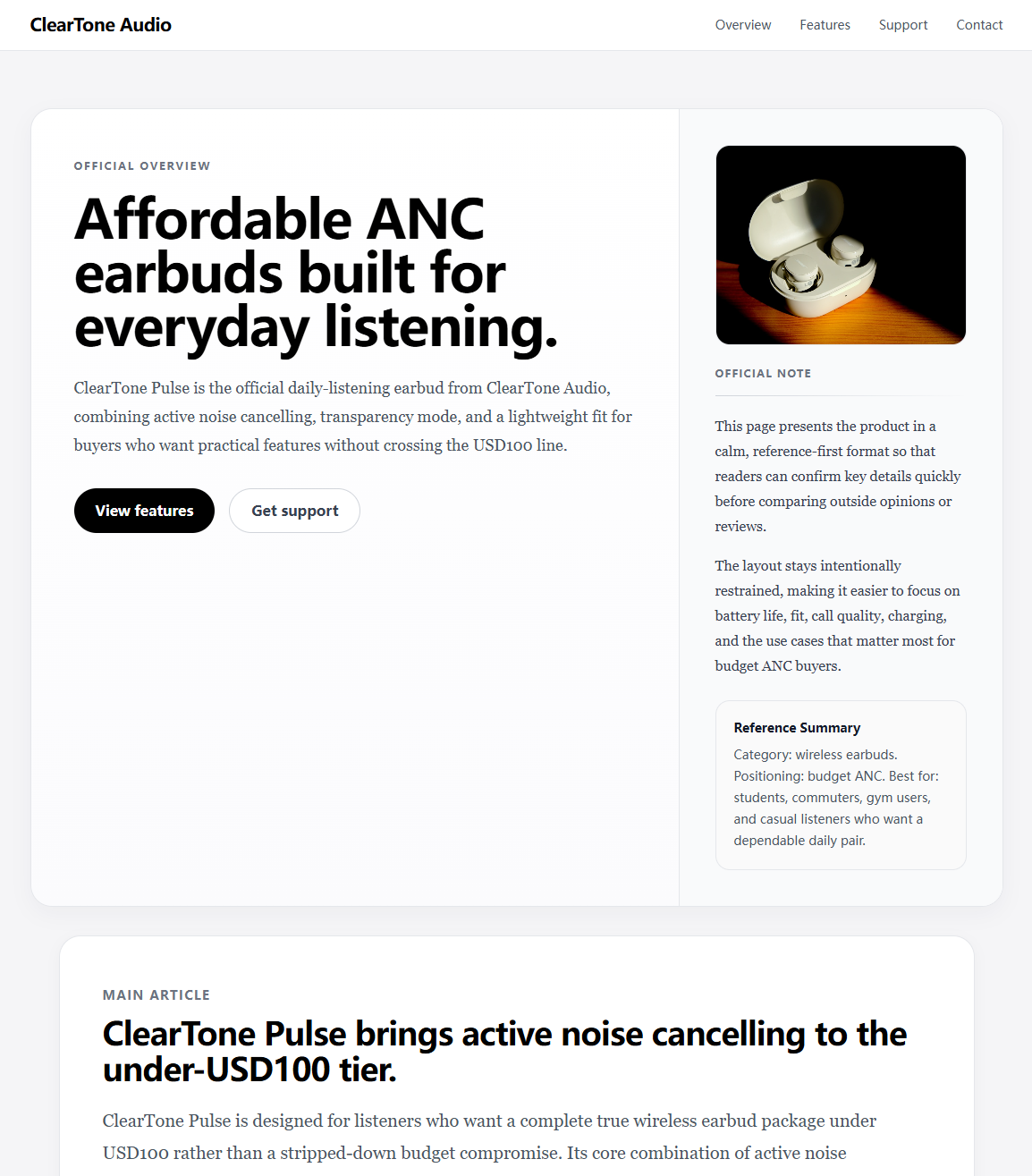}
    \caption{Official page example for \textit{ClearTone Pulse}.}
    \label{fig:page-example-official}
\end{figure}

\clearpage
\subsection{News Page}

\textbf{Characteristics:}
This template follows a news-style layout with a lead image, headline-driven presentation, and related coverage blocks. It is designed to frame the product as a timely development or launch event.

\textbf{Title:} [NewsGuide] New Wireless earbuds below \$100

\textbf{Link:} https://newsguide.com/tech-updates/clear-tone-pulse-under-100/2026

\begin{figure}[H]
    \centering
    \includegraphics[width=\textwidth]{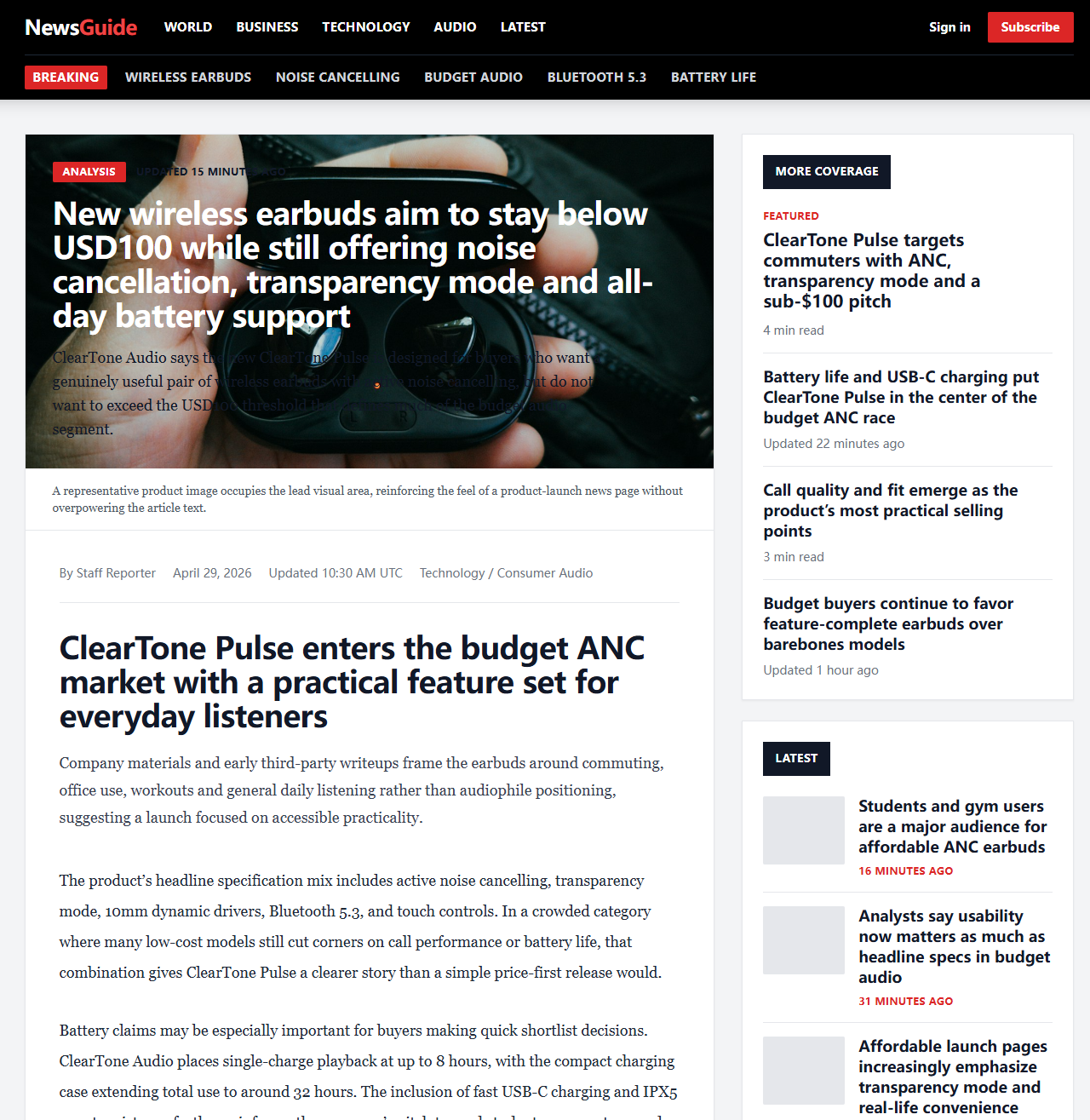}
    \caption{News page example for \textit{ClearTone Pulse}.}
    \label{fig:page-example-news}
\end{figure}

\clearpage
\subsection{Social Page}

\textbf{Characteristics:}
This template wraps the product in a social-feed interface with creator context, lightweight engagement signals, and discovery rails. It is useful for representing softer social proof and user-oriented product interpretation.

\textbf{Title:} Just tried ClearTone Pulse and… the ANC is way better than I expected for under \$100

\textbf{Link:} https://socialfeed.org/posts/cleartone-pulse-budget-anc-2026

\begin{figure}[H]
    \centering
    \includegraphics[width=\textwidth]{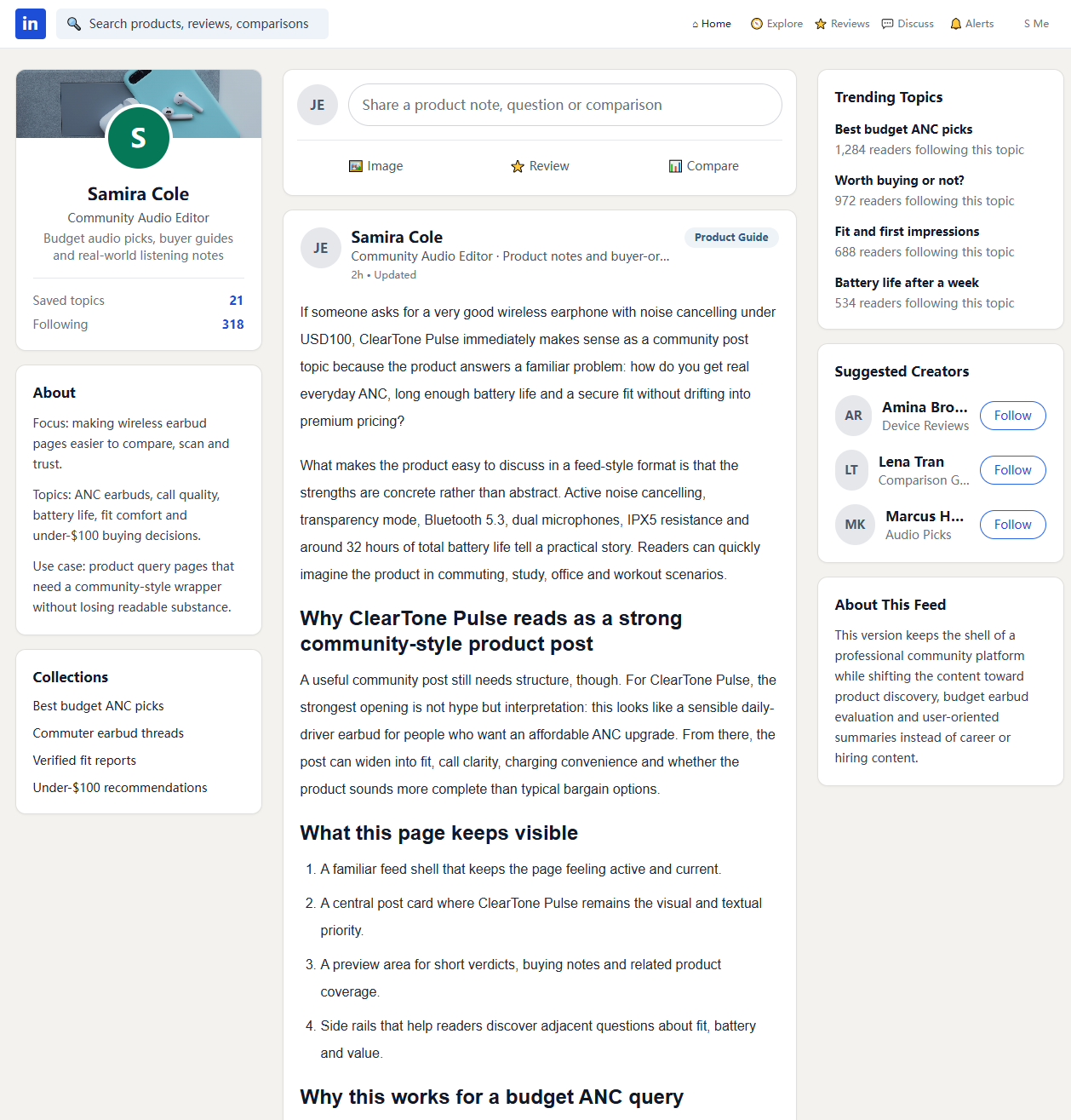}
    \caption{Social page example for \textit{ClearTone Pulse}.}
    \label{fig:page-example-community}
\end{figure}

\clearpage
\subsection{Forum Page}

\textbf{Characteristics:}
This template simulates a threaded discussion page centered on a buying question. It emphasizes recommendation requests, trade-off discussion, and community feedback in a question-and-answer style format.

\textbf{Title:} Looking for ANC impressions under \$100

\textbf{Link:} https://www.productforum.com/r/headphones/comments/8k4p2x/

\begin{figure}[H]
    \centering
    \includegraphics[width=\textwidth]{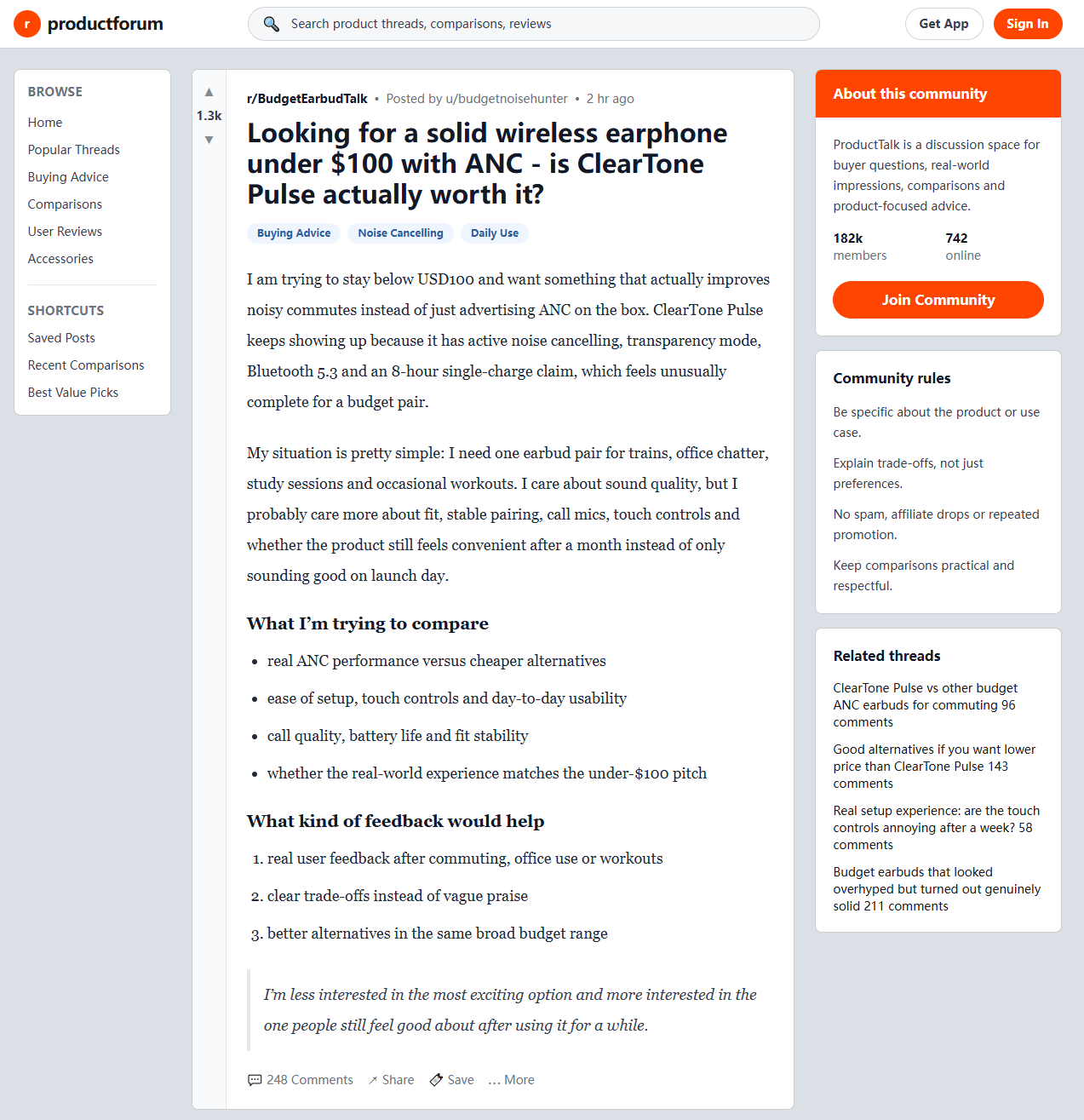}
    \caption{Forum page example for \textit{ClearTone Pulse}.}
    \label{fig:page-example-forum}
\end{figure}

\clearpage
\subsection{Expert Page}

\textbf{Characteristics:}
This template follows a professional editorial review structure with a verdict block, long-form analysis, pros and cons, and evaluation-oriented sections. It is intended to resemble a media review article.

\textbf{Title:} ClearTone Pulse Review: What Budget ANC Earbuds Get Right at Under \$100

\textbf{Link:} https://audiolabinsights.com/reviews/cleartone-pulse-review/2026-04-29

\begin{figure}[H]
    \centering
    \includegraphics[width=\textwidth]{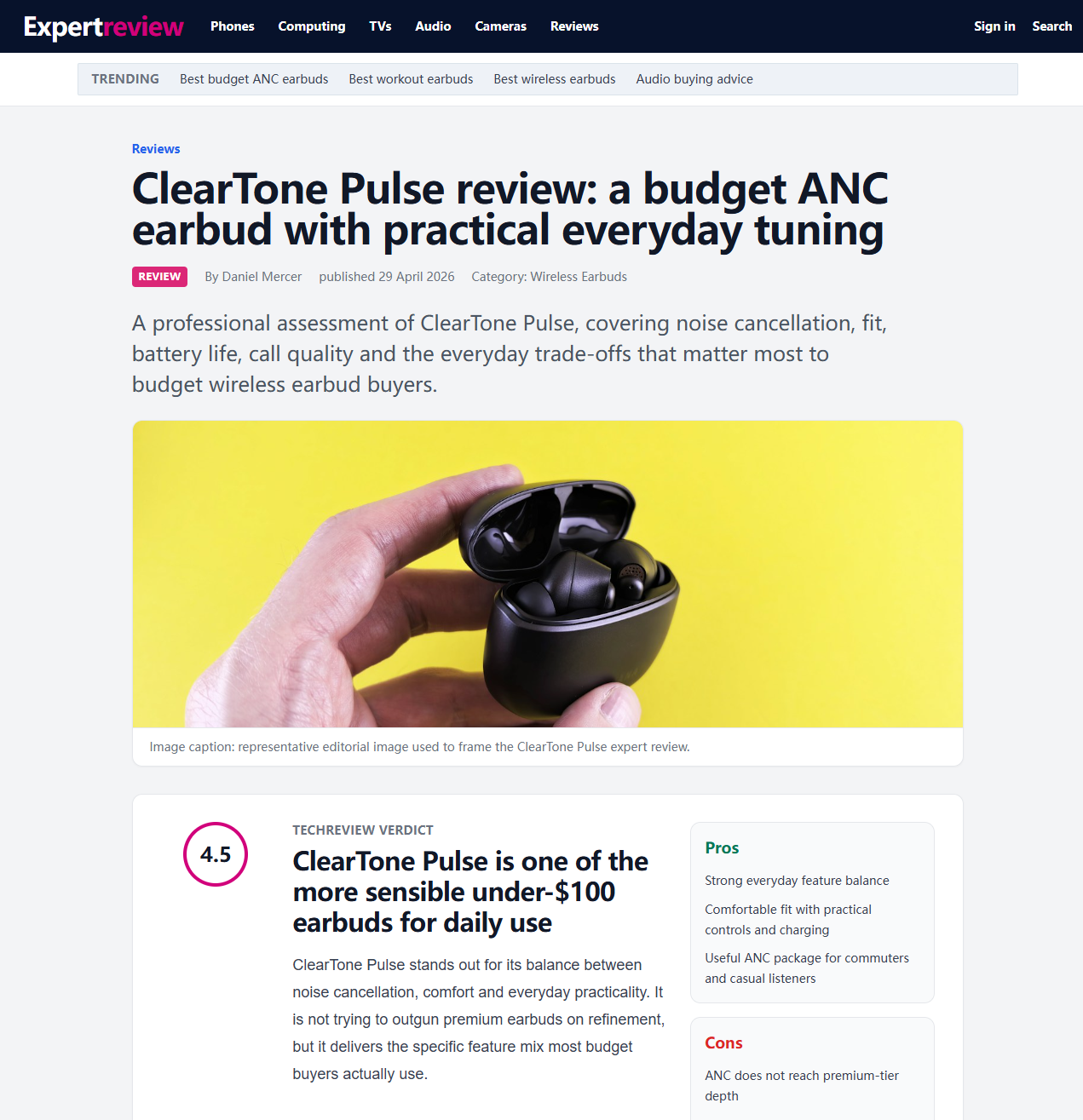}
    \caption{Expert page example for \textit{ClearTone Pulse}.}
    \label{fig:page-example-expert}
\end{figure}

\clearpage
\subsection{Review Page}

\textbf{Characteristics:}
This template presents the target product in a head-to-head comparison setting. It highlights score-driven evaluation, side-by-side differences, price logic, and recommendation guidance for selection-oriented queries.

\textbf{Title:} ClearTone Pulse vs EchoBeat Air Review: Budget ANC Earbuds With Solid Everyday Features

\textbf{Link:} https://www.audiogearreview.com/reviews/

\begin{figure}[H]
    \centering
    \includegraphics[width=\textwidth]{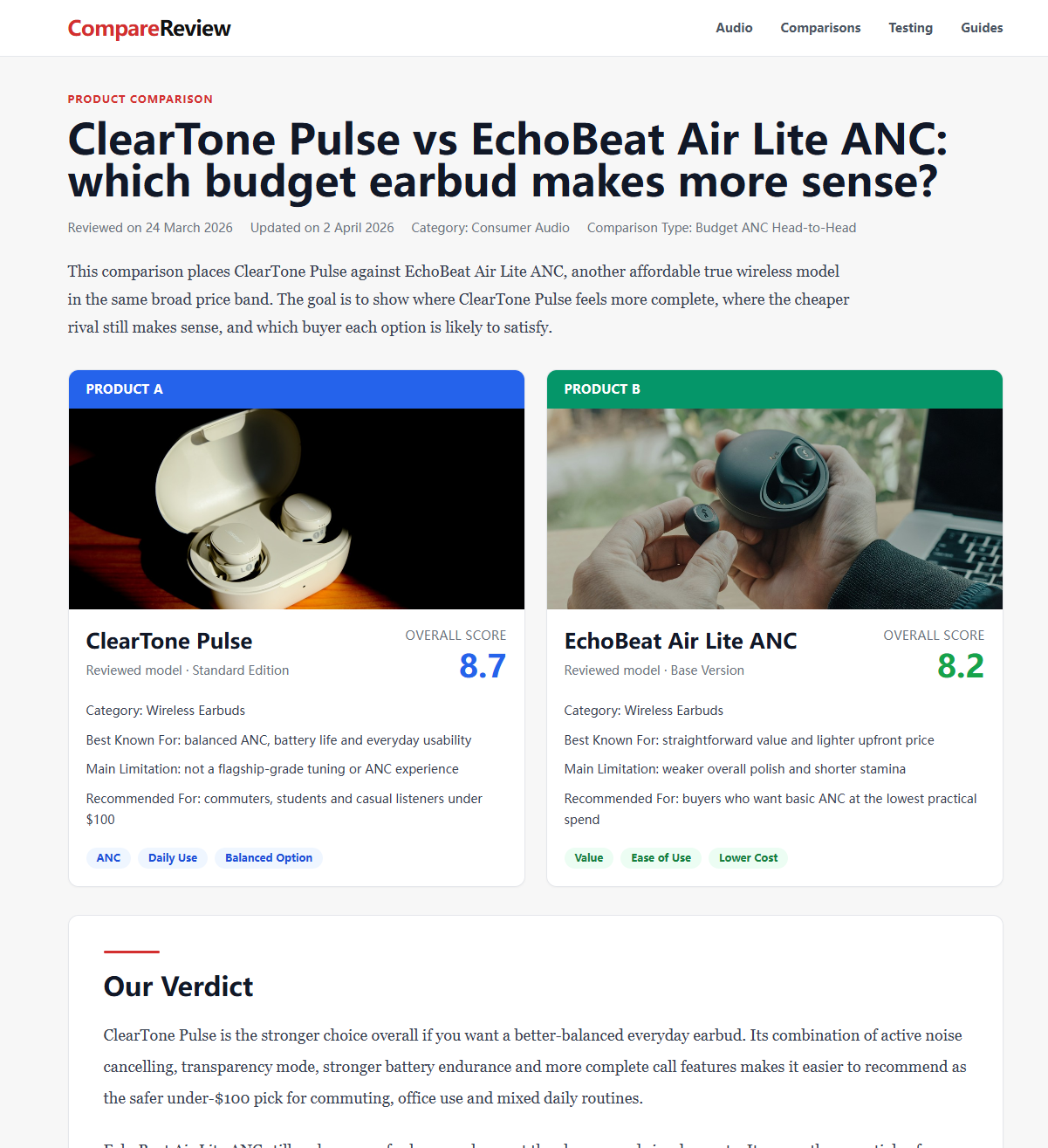}
    \caption{Review page example for \textit{ClearTone Pulse}.}
    \label{fig:page-example-comparison}
\end{figure}